\documentclass[sigconf,natbib=true]{acmart}
\usepackage{kotex}
\usepackage{placeins}
\usepackage{float}
\usepackage{microtype}
\usepackage{enumitem}
\usepackage{booktabs}
\usepackage{array}
\usepackage{multirow}
\usepackage{tabularx}
\usepackage{subcaption}
\usepackage{balance}
\usepackage{hyperref}
\usepackage{pgfplots}
\usetikzlibrary{patterns,patterns.meta}
\pgfplotsset{compat=1.18}
\usepackage[normalem]{ulem}
\usepackage{pgfplotstable}
\usepackage{mathtools}
\useunder{\uline}{\ul}{}
\newcolumntype{C}{>{\centering\arraybackslash}p{2cm}}
\hypersetup{
    colorlinks=true,
    linkcolor=blue,
    urlcolor=cyan
}
\usepackage{tikz}
\usepackage{pgfplots}
\pgfplotsset{compat=1.18}



\definecolor{darkblue}{rgb}{0.0, 0.0, 0.55}
\definecolor{ao(english)}{rgb}{0.0, 0.5, 0.0}
\definecolor{coolblack}{rgb}{0.0, 0.18, 0.39}
\definecolor{purpleheart}{rgb}{0.41, 0.21, 0.61}
\definecolor{pastelviolet}{rgb}{0.8, 0.6, 0.79}
\definecolor{lightskyblue}{rgb}{0.53, 0.81, 0.98}
\definecolor{palecornflowerblue}{rgb}{0.67, 0.8, 0.94}
\definecolor{lightmauve}{rgb}{0.86, 0.82, 1.0}
\definecolor{lightpastelpurple}{rgb}{0.69, 0.61, 0.85}
\definecolor{bronze}{rgb}{0.8, 0.5, 0.2}
\definecolor{armygreen}{rgb}{0.29, 0.33, 0.13}
\definecolor{darkpowderblue}{rgb}{0.0, 0.2, 0.6}
\definecolor{falured}{rgb}{0.5, 0.09, 0.09}
\definecolor{outerspace}{rgb}{0.25, 0.29, 0.3}
\definecolor{tangerine}{rgb}{0.95, 0.52, 0.0}
\definecolor{seagreen}{rgb}{0.18, 0.55, 0.34}
\definecolor{springgreen}{rgb}{0.0, 1.0, 0.5}
\definecolor{applegreen}{rgb}{0.55,0.71,0.0}
\definecolor{amethyst}{rgb}{0.6,0.4,0.8}
\definecolor{amber}{rgb}{1.0,0.49,0.0}
\definecolor{darkgreen}{rgb}{0,0.4,0} 

\usepackage[table]{xcolor}
\usepackage{pgfplots}
\usetikzlibrary{plotmarks}
\usepgfplotslibrary{groupplots}

\AtBeginDocument{%
  }

\begin{document}

\title{Anchored Alignment: Preventing Positional Collapse \\ in Multimodal Recommender Systems}
% }

\author{Yonghun Jeong}
\affiliation{
	\institution{UNIST}
  	\country{Ulsan, Korea}
}
\email{younghune135@gmail.com}

\author{David Yoon Suk Kang}
\affiliation{
	\institution{Chungbuk National University}
  	\country{Cheongju, Korea}
}
\email{dyskang@cbnu.ac.kr}

\author{Yeon-Chang Lee}
\authornote{Corresponding author}
\affiliation{
	\institution{UNIST}
  	\country{Ulsan, Korea}
}
\email{yeonchang@unist.ac.kr}

%\renewcommand{\shortauthors}{}

\begin{abstract}

Multimodal recommender systems (MMRS) leverage images, text, and interaction signals to enrich item representations. 
However, recent alignment-based MMRSs that enforce a unified embedding space often blur modality-specific structures and exacerbate ID dominance.
Therefore, we propose \ours, a multimodal recommendation framework that performs indirect, anchor-based alignment in a lightweight projection domain.
By decoupling alignment from representation learning, \ours\ preserves each modality’s native structure while maintaining cross-modal consistency and avoiding positional collapse.
Experiments on four Amazon datasets show that \ours\ achieves competitive top-$N$ recommendation accuracy, while qualitative analyses demonstrate improved multimodal expressiveness and coherence.
The codebase of \ours\ is available at
\url{https://github.com/hun9008/AnchorRec}.
\end{abstract}
% <concept_id>10002951.10003227.10003351.10003269</concept_id>
% <concept_desc>Information systems~Collaborative filtering</concept_desc>

%\ccsdesc[500]{Information systems~Collaborative filtering}

%\keywords{review-based recommendation system, user likes and dislikes, homogeneous graph}
\newcommand{\spec}{{\it spec.}}
\newcommand{\aka}{{\it a.k.a.}}
\newcommand{\ie}{{\it i.e.}}
\newcommand{\eg}{{\it e.g.}}
\newcommand{\ours}{\textsc{\textsf{AnchorRec}}}
\newcommand{\blue}{\textcolor{blue}}

\newcommand{\REGfull}{anchor modal preserve loss}
\newcommand{\UIAfull}{user item margin loss}
\newcommand{\REGfullname}{Anchor Modal Preserve Loss}
\newcommand{\UIAfullname}{User Item Margin Loss}
\newcommand{\REG}{AMP}
\newcommand{\UIA}{UIM}

\newcommand{\mj}[1]{\textcolor{blue}{[MJ: #1]}}
\newcommand{\jw}[1]{\textcolor{green}{[JW: #1]}}
\newcommand{\yc}[1]{\textcolor{red}{[YC: #1]}}

%eq2
\newcommand{\oursall}{\textsc{\textsf{{TraceRec(all)}}}}
\newcommand{\oursforward}{\textsc{\textsf{{TraceRec(forward)}}}}
\newcommand{\oursrecent}{\textsc{\textsf{{TraceRec(recent)}}}}

%eq3
\newcommand{\ourswoproj}{\textsc{\textsf{{TraceRec(w/o proj)}}}}

\setcopyright{none}
\settopmatter{printacmref=false}
\renewcommand\footnotetextcopyrightpermission[1]{}
\pagestyle{plain}
\maketitle
\section{Introduction} \label{sec:intro} \begin{figure}[t]
\centering
\includegraphics[width=1\linewidth]{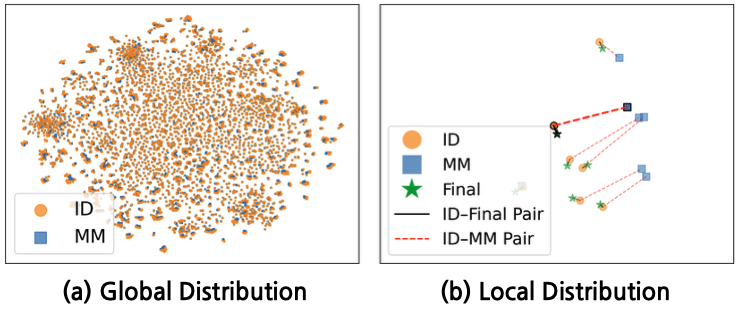}
% \vspace{-0.7cm}
% \caption{
% t-SNE visualization of item embeddings from AlignRec~\cite{liu2024alignrec}. 
% ID embeddings (blue) and modality-specific embeddings (orange) collapse tightly into the final aligned embeddings (green). 
% Red lines highlight how different modalities of the same item are pulled into a single point.} \label{fig:intro}

%% modify
% \caption{t-SNE~\cite{maaten2008visualizing} visualization of AlignRec~\cite{liu2024alignrec}. 
% % (left) and AnchorRec (right).
% The top panel shows the global distribution of ID (orange) and modality-specific (blue) embeddings, while the bottom panel presents a zoomed-in view of selected items.
\caption{t-SNE~\cite{maaten2008visualizing} visualization of embeddings from AlignRec~\cite{liu2024alignrec}:
the left panel shows the global distribution, and the right panel provides a zoomed-in local view revealing fine-grained local structure.
% In AlignRec, ID embeddings (orange) and modality-specific embeddings (blue) are closely aligned (C1),
% and the zoomed view shows final item embeddings (green) being dominated by ID embeddings, as indicated by red lines (C2).
} \label{fig:intro}
%% modify
\vspace{-0.5cm}
\end{figure}

\noindent\textbf{Background.} Recommender systems play a crucial role in e-comm- erce and content platforms by identifying items that match user preferences~\cite{rendle2012bpr, he2020lightgcn}.
As preferences are shaped by heterogeneous signals, such as images, textual descriptions, and behavioral interactions, \textbf{multimodal recommender systems} \textbf{(MMRSs)}~\cite{he2016vbpr, zhang2021mining, zhou2023tale, zhou2023bootstrap, xv2024improving, guo2024lgmrec, ong2025spectrum, liu2024alignrec, wu2025aligning, kim2022mario, zhou2023enhancing, ma2019multi, yu2025mind, zhong2024mirror, ferrari2019we, yi2021multi, wang2021dualgnn, tao2022self, wei2019mmgcn, wei2020graph} have emerged as a key direction for integrating heterogeneous signals into recommendation models.

Early MMRSs, including VBPR~\cite{he2016vbpr} and FREEDOM~\cite{zhou2023tale}, incorporated multimodal features into ID-based recommendation frameworks to alleviate data sparsity and cold-start issues.
Despite their effectiveness, these models treat multimodal signals largely as auxiliary features, combining them through simple fusion mechanisms.
Because each modality encodes user preferences in a heterogeneous feature space, such fusion induces only weak cross-modal alignment, leaving discrepancies across modalities insufficiently resolved.
% As a consequence, they struggle to reconcile cross-modal discrepancies and to fully exploit the complementary semantics encoded in different modalities.
Motivated by this limitation, recent alignment-based methods, including DA-MRS~\cite{xv2024improving} and AlignRec~\cite{liu2024alignrec}, explicitly project all modalities into a \textit{shared latent space}, typically using contrastive learning or distance-minimization objectives.

\vspace{1mm}
\noindent\textbf{Challenges.}
While projection-based alignment improves consistency, it introduces a fundamental trade-off (\textbf{Figure~\ref{fig:intro}}): \textbf{forcing all modalities to converge into a single space reduces modality-specific expressiveness}.
This trade-off leads to two key challenges that limit the representational capacity of existing MMRSs.

\textbf{(C1) Positional Collapse of Modality Representations.} 
Most alignment-based MMRSs encourage embeddings from different modalities (\eg, image, text, and ID) of the same item to become as close as possible in a shared latent space.
%% modify
As shown in \textbf{Figure~\ref{fig:intro}-(a)}, AlignRec~\cite{liu2024alignrec}, a representative alignment-based method, collapses modality-specific embeddings of an item into nearly identical positions.
% from different modalities of the same item into a single point in the shared space.
As a result, distinct semantic signals across modalities are compressed into a single point, reducing semantic diversity and diminishing modality-specific characteristics.
% Such direct alignment methods often lead all modality embeddings to collapse into a single point, reducing semantic diversity and erasing modality-specific characteristics.
%% modify

% This often drives all modality embeddings to collapse into a single point, reducing semantic diversity and erasing modality-specific characteristics.

\textbf{(C2) Overdominance of Interaction Signals.}
Beyond alignment itself, most MMRSs further optimize item representations using interaction-driven objectives such as the \textit{Bayesian Personalized Ranking} (BPR)~\cite{rendle2012bpr}.
%% modify
As shown in \textbf{Figure~\ref{fig:intro}-(b)}, 
this optimization causes the final item embeddings to be heavily biased toward ID-based interaction patterns, even after multimodal fusion.
Consequently, interaction signals dominate the learned representations, suppressing multimodal semantics and reinforcing the positional collapse observed in (C1). 
%% modify
% In contrast, as shown in the bottom panel of Figure~\ref{fig:intro}-(b), \ours\ yields representations that are not dominated by any single embedding signal.
%% modify

\vspace{1mm}
\noindent\textbf{Proposed Ideas.}
To address the above challenges, we aim to enhance the expressiveness of multimodal item representations while maintaining competitive predictive accuracy.
To this end, we propose \textbf{\ours}, an \textit{anchor-based alignment} framework that mitigates positional collapse induced by direct alignment and achieves cross-modal consistency via an intermediate anchoring space.

\begin{figure*}[t]
\centering
\includegraphics[width=0.99\linewidth]{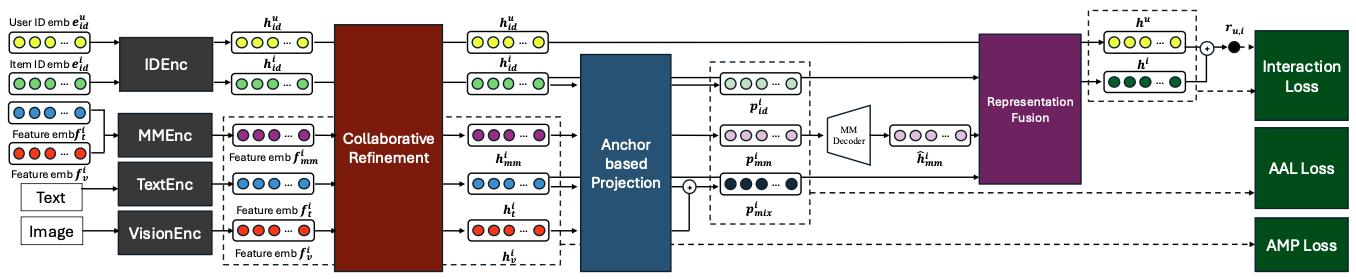}
\vspace{-0.4cm}
\caption{Overview of \ours, highlighting the \textbf{anchor-based projection} and its alignment losses as the core mechanism.
% , consisting of modality encoders, collaborative refinement, anchor-based projection, and representation fusion.
}
\label{fig:overview}
\vspace{-0.2cm}
\end{figure*}

Unlike prior methods that force all modalities to overlap within a single latent space, \ours\ introduces a lightweight \textbf{projection domain} in which modalities are aligned indirectly.
Within this domain, we design an \textbf{anchor-based alignment loss} that uses a fused multimodal embedding as a \textit{shared anchor}.
The anchor provides a stable semantic reference that guides the projected ID, text, and vision representations toward semantic agreement, \textit{while preserving their modality-specific structure in the original spaces}.

\vspace{1mm}
\noindent\textbf{Contributions.} Our contributions are summarized as follows:
\begin{itemize}[leftmargin=*]
    \item \textbf{Indirect Alignment Strategy:} We propose a projection-domain alignment strategy that aligns modalities \textit{indirectly}, alleviating positional collapse while preserving modality-specific structure.
    \item \textbf{Anchor-based Alignment Loss:} We design an anchor-guided loss that uses a fused multimodal embedding as a shared reference, reducing ID dominance and promoting balanced semantic agreement across ID, text, and vision modalities.
    \item \textbf{Empirical Validation of Multimodal Expressiveness:} \linebreak Through experiments on four real-world datasets, we demonstrate that \ours\ enhances the expressiveness of multimodal representations while preserving competitive recommendation accuracy.
\end{itemize}
% \section{Introduction} \label{sec:intro} \input{Sections/short/Intro_short}

\section{Related Work} \label{sec:rworks} \noindent\textbf{Multimodal Recommender Systems.}
Early MMRSs, such as VBPR~\cite{he2016vbpr}, integrated visual features into collaborative filtering to enrich item representations.
Subsequent models (\eg, LATTICE~\cite{zhang2021mining} and FREEDOM~\cite{zhou2023tale}) exploited multimodal relations or auxiliary losses to enhance ID-based representations. 
More recent architectures (\eg, SMORE~\cite{ong2025spectrum}, LGMRec~\cite{guo2024lgmrec}) focus on improving multimodal fusion to better alleviate data sparsity and cold-start issues. 
Despite their effectiveness, these approaches primarily treat multimodal features as auxiliary side information and do not fundamentally address the representational gap across modalities.

\noindent\textbf{Multimodal Alignment for Recommendation.}
To explicitly reduce cross-modal inconsistency, alignment-based MMRSs project modalities into a unified latent space.
% using contrastive objectives (BM3~\cite{zhou2023bootstrap}), noise-resistant alignment (DA-MRS~\cite{xv2024improving}), or multi-level alignment losses (AlignRec~\cite{liu2024alignrec}). 
Representative approaches employ contrastive learning objectives, as in BM3~\cite{zhou2023bootstrap}, noise-robust alignment strategies, as in DA-MRS~\cite{xv2024improving}, or multi-level alignment losses, as in AlignRec~\cite{liu2024alignrec}.
%% remove
% Techniques such as Wasserstein-based distribution alignment and gradient modulation (AB-Rec~\cite{wu2025aligning}) further enhance consistency between ID and multimodal embeddings. 
%% remove
While these methods improve cross-modal consistency, they rely on direct alignment in a shared space, which can limit modality-specific characteristics. 
% \section{Related Work} \label{sec:rworks} \input{Sections/short/RWorks_short}

% \section{Proposed Framework: \ours} \label{sec:method} \input{Sections/Method}
\section{The Proposed Framework: \ours} \label{sec:method} \subsection{Problem Definition}\label{sec:method-overview} 
Let $\mathcal{U}$ and $\mathcal{I}$ denote the sets of users and items, respectively.
User–it\ em interactions are represented as a bipartite graph $\mathcal{G} = (\mathcal{V}, \mathcal{E})$, 
% where $\mathcal{V}=\mathcal{U}\cup\mathcal{I}$, $\mathcal{E}=\{(u,i)|r_{u,i}=1\}$, with $r_{u,i}=1$ indicating that user $u$ has interacted with item $i$.
where $(u,i)\in\mathcal{E}$ indicates that user $u$ has interacted with item $i$.
We denote $\mathcal{N}_u$ the set of items interacted with by user $u$.
Each user $u$ and item $i$ is associated with a $d$-dimensional ID embedding, $\mathbf{h}_{id}^u, \mathbf{h}_{id}^i \in \mathbb{R}^d$.
In addition, each item $i$ has modality-specific feature embeddings extracted from pretrained encoders, including multimodal $\mathbf{f}_{mm}^i,$ textual $\mathbf{f}_{t}^i,$ and visual $\mathbf{f}_{v}^i$ representations.
Accordingly, the set of item modalities is $\mathcal{M} = \{\mathrm{id},\ \mathrm{mm},\ \mathrm{t},\ \mathrm{v}\}$.
Given a user $u$, the task is to rank items in $\mathcal{I} \setminus \mathcal{N}_u$ and recommend the top-$k$ items.
Beyond accuracy, we further require the model to integrate multimodal signals while preserving modality-specific representational structures.

\subsection{Key Components}\label{sec:method-module} 
\textbf{Figure~\ref{fig:overview}} provides an overview of \ours, which consists of four components: (1) modality encoders, (2) collaborative refinement, (3) anchor-based projection, and (4) representation fusion.

\vspace{1mm}
\noindent\textbf{Modality Encoders.}  
\ours\ initializes each user $u$ and item $i$ with ID embeddings produced by an ID-based encoder $\mathrm{IDEnc}(u,i)$, yielding $\mathbf{h}_{id}^u$ and $\mathbf{h}_{id}^i$.
Any CF-based encoder can be adopted as $\mathrm{IDEnc}$, and we use LightGCN ~\cite{he2020lightgcn} as the backbone.
For each item $i$, pretrained modality encoders extract modality-specific feature embeddings: 
(1) $\mathrm{TextEnc}(i)$ (\eg, BERT~\cite{devlin2019bert}) for textual features $\mathbf{f}_t^i$, 
(2) $\mathrm{VisionEnc}(i)$ (\eg, VGGNet~\cite{simonyan2014very}) for image features $\mathbf{f}_v^i$, and 
(3) $\mathrm{MMEnc}(i)$ (\eg, BEiT3~\cite{wang2023image}) for multimodal fusion features $\mathbf{f}_{mm}^i$.
Here, $\mathrm{MMEnc}$ jointly encodes textual and visual inputs, $\mathbf{f}_t^i$, $\mathbf{f}_v^i$.\footnote{$\mathrm{MMEnc}$ follows the multimodal encoder design of AlignRec~\cite{liu2024alignrec}, where image patch tokens and text tokens are modeled within a unified token sequence with a shared \textbf{[CLS]} token.
During fine-tuning, masked reconstruction objectives are applied separately to visual and textual tokens.}
%% modify
All modality-specific feature embeddings are fixed during training and serve as input signals for subsequent modules.

\vspace{1mm}
\noindent\textbf{Collaborative Refinement.} 
Item-side modality features provide rich semantic information but do not capture  collaborative signals from user–item interactions.
In contrast, users do not have inherent modality features and must express their preferences through the items they have consumed.
Accordingly, \ours\ bridges this asymmetry by injecting interaction-driven information into item modality embeddings and by constructing user-side modality preferences from interacted items.
For \textit{item-side refinement}, each modality feature $\mathbf{f}_m^i$ is first modulated by the corresponding item ID embedding $\mathbf{h}_{\mathrm{id}}^{i}$, yielding $\widetilde{\mathbf{h}}_{m}^{i} = \mathbf{h}_{\mathrm{id}}^{i} \odot \sigma(\mathrm{MLP}(\mathbf{f}_m^i))$.
The refined embeddings are then propagated across similar items using an item-item similarity matrix $\mathcal{S}\in \mathbb{R}^{|\mathcal{I}|\times|\mathcal{I}|}$~\cite{liu2024alignrec}, yielding $\mathbf{h}_{m}^{i} = \sum_{j \in \mathcal{I}} \mathcal{S}_{ij} \widetilde{\mathbf{h}}_{m}^{i}$.

\vspace{1mm}
\noindent\textbf{Anchor-based Projection.}
Directly aligning modalities in a shared latent space often causes positional collapse and degrades modality-specific structure.
To avoid this, \ours\ performs alignment only within a lightweight \textit{projection domain} and \textit{only on item embeddings}, while preserving the original modality space.
Specifically, each item-side modality embedding is mapped into the projection domain via a small MLP, \ie, $\mathbf{p}_{\mathrm{m}}^{i} = \mathrm{MLP}_m(\mathbf{h}_{\mathrm{m}}^{i})$ for $m\in\{\mathrm{id}, \mathrm{mm}, \mathrm{t}, \mathrm{v}\}$.
Within this domain, the multimodal projection $\mathbf{p}_{\mathrm{mm}}^{i}$ serves as a shared \textit{anchor}, and the ID, textual, and visual projections ($\mathbf{p}_{\mathrm{id}}^{i}$, $\mathbf{p}_{\mathrm{t}}^{i}$, and $\mathbf{p}_{\mathrm{v}}^{i}$)
are softly pulled toward this anchor using an anchor-based alignment loss.
We restrict projection to items since users lack modality features, and projecting user embeddings would add unnecessary instability without benefiting cross-modal alignment.

\vspace{1mm}
\noindent\textbf{Representation Fusion.}
\ours\ constructs the final user and item representations by fusing the signals from ID, multimodal, textual, and visual embeddings.
Specifically, the item representation is obtained as $\mathbf{h}^{i}=\sum_{\mathrm{m} \in \mathcal{M}} \mathbf{h}_{\mathrm{m}}^{i}$, while the user representation is given by $\mathbf{h}^{u}=\mathbf{h}_{\mathrm{id}}^{u}$.
To further incorporate multimodal semantics, we inject a correction term derived from the projected multimodal anchor, $\mathbf{h}^{i}=\mathbf{h}^{i}+\lambda_{\mathrm{recon}}\mathrm{MLP}(\mathbf{p}_{\mathrm{mm}}^{i})$, where $\lambda_{\mathrm{recon}}$ controls the contribution of this correction. 
This term gently adjusts item embeddings toward the multimodal anchor without disrupting modality-specific structure. 
As a result, the final representation better captures multimodal information and avoids being dominated by ID signals.
The resulting representations $\mathbf{h}^u$ and $\mathbf{h}^i$ are then used to compute the final relevance score for recommendation.

\subsection{Training}\label{sec:method-module} 

\begin{figure}[t]
\centering
\includegraphics[width=\linewidth]{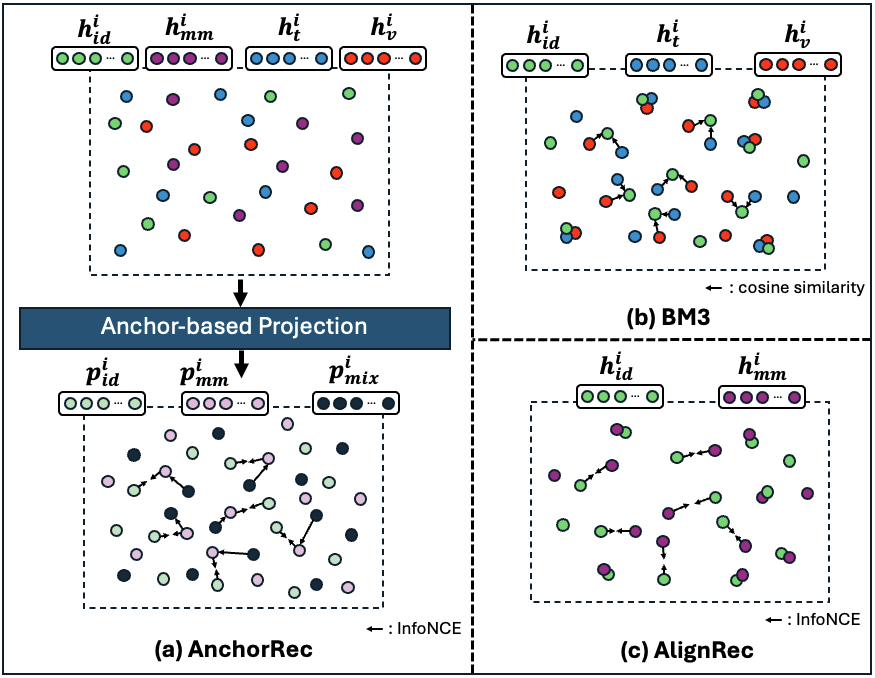}
\vspace{-0.2cm}
% \caption{Illustration of anchor alignment in the projection domain.
% The multimodal projection (anchor, $\mathbf{p}_{mm}^i$) pulls ID, text, and vision projections toward itself.
% } \label{fig:alignment}
%% modify
\caption{Comparison of alignment strategies:
(a) \ours: anchor-based \textit{indirect} alignment via a shared multimodal anchor ($\boldsymbol{\mathbf{p}_{\mathrm{mm}}^{i}}$);
(b) BM3~\cite{zhou2023bootstrap}: \textit{direct} similarity-based alignment across modalities; and
(c) AlignRec~\cite{liu2024alignrec}: \textit{direct} contrastive alignment between ID and multimodal embeddings.
} \label{fig:alignment}
%% modify
\vspace{-0.5cm}
\end{figure}

\ours\ is trained by jointly optimizing interaction ranking and projection-domain alignment.
The overall objective is defined as: 

\vspace{-0.2cm}
\footnotesize
\begin{equation}
\mathcal{L}
= \mathcal{L}_{\mathrm{interaction}}
% + \lambda_{1}\, \mathcal{L}_{\mathrm{\REG}}
% + \mathcal{L}_{\mathrm{\UIA}}
+ \mathcal{L}_{\mathrm{anchor}}.
\end{equation}
\normalsize 
The first term, $\mathcal{L}_{\mathrm{interaction}}$, follows prior alignment-based MMRS methods~\cite{liu2024alignrec} and optimizes interaction-aware ranking.

Due to space limitations, we omit their formulations and focus on our key contribution, the \textbf{anchor-based alignment loss} $\mathcal{L}_{\mathrm{anchor}}$, which enables stable cross-modal alignment without collapsing modality-specific structure.
Motivated by the observations in Section~\ref{sec:intro} that interaction-driven objectives can induce positional collapse and amplify ID dominance, we design  $\mathcal{L}_{\mathrm{anchor}}$ to regulate alignment in a controlled manner.
Specifically, $\mathcal{L}_{\mathrm{anchor}}$ consists of two complementary terms: $\mathcal{L}_{\mathrm{anchor}}
= \lambda_{1} \mathcal{L}_{\mathrm{AAL}} + \lambda_{2} \mathcal{L}_{\mathrm{AMP}}$, where $\lambda_{1}$ and $\lambda_{2}$ control their relative contributions.

\textbf{Anchor-Aligned Learning (AAL)} softly aligns ID and single-modality projections with a multimodal anchor in the projection domain using an InfoNCE~\cite{he2020momentum, zhao2023augmented} objective as follows:
% We implement AAL using an InfoNCE objective that pulls paired projections toward the anchor while separating unpaired samples:

\vspace{-0.2cm}
\footnotesize
\begin{equation}
\mathcal{L}_{\mathrm{AAL}} = \mathrm{InfoNCE}(\mathbf{p}_{\mathrm{mm}}^{i},\, \mathbf{p}_{\mathrm{id}}^{i})
+ \mathrm{InfoNCE}(\mathbf{p}_{\mathrm{mm}}^{i},\, \mathbf{p}_{\mathrm{mix}}^{i}),
\end{equation}
\normalsize
where $\mathbf{p}_{\mathrm{mix}}^{i}=\alpha\, \mathbf{p}_{\mathrm{t}}^{i} + (1-\alpha)\, \mathbf{p}_{\mathrm{v}}^{i} $ combines text and vision projections.
%% modify
\textbf{Figure~\ref{fig:alignment}} highlights the key differences between AAL and existing alignment strategies.
Unlike BM3~\cite{zhou2023bootstrap} (\textbf{Figure~\ref{fig:alignment}-(b)}) and AlignRec~\cite{liu2024alignrec} (\textbf{Figure~\ref{fig:alignment}-(c)}), which rely on direct alignment and often suffer from positional collapse and ID dominance, AAL addresses both issues through two key design choices (\textbf{Figure~\ref{fig:alignment}-(a)}).
First, AAL uses the multimodal embedding as a shared anchor instead of ID embeddings, reducing interaction-driven ID dominance.
Second, AAL performs \textit{indirect} alignment in a projection domain rather than direct alignment in the original space, decoupling alignment from representation learning and alleviating positional collapse.
% which treats ID embeddings as alignment targets, AAL uses the multimodal embedding as the anchor.
% This choice prevents the learned representations from being dominated by ID-based interaction signals, as illustrated in Figure~\ref{fig:alignment}-(a) and (b).
% In addition, unlike AlignRec~\cite{liu2024alignrec}, which  directly aligns modality embeddings in the original embedding space, AAL adopts an \textit{indirect} alignment strategy in a projection domain.
% By decoupling alignment from representation learning, this design alleviates positional collapse, as illustrated in Figure~\ref{fig:alignment}-(a) and (c).
% Moreover, 
%% modify

\textbf{Anchor-Modality Preservation (AMP)} preserves each modality’s intrinsic similarity structure from pretrained features by penalizing deviations in learned similarities:

\vspace{-0.2cm}
\footnotesize
\begin{equation}
\mathcal{L}_{\mathrm{AMP}} = \sum_{m \in \{\mathrm{mm}, \mathrm{t}, \mathrm{v}\}} 
   \lambda_{m} \left\|\, 
   C(\mathbf{h}_{m}^{i}, \mathbf{h}_{m}^{j}) - C(\mathbf{f}_{m}^{i}, \mathbf{f}_{m}^{j}) 
   \,\right\|_{2},
\end{equation}
\normalsize
where $C(\cdot,\cdot)$ denotes cosine similarity. 
Together, AAL enforces anchor-guided alignment in the projection domain, while AMP preserves modality-specific structure, resulting in expressive yet stable multimodal representations.

\section{Evaluation} \label{sec:eval} % \begin{table}[t]
% \footnotesize
% \centering
% \caption{Dataset statistics}
% \vspace{-0.2cm}
% \label{Table:dataset}
% \resizebox{.48\textwidth}{!}{
%  \renewcommand{\arraystretch}{1}
% \begin{tabular}{crrrrr}
% \toprule
% \textbf{Datasets} & \textbf{\# users} & \textbf{\# items} & \textbf{\# interactions} & \textbf{\# timestamps} \\ \midrule
% \textbf{LastFM} & 980  & 1,000  & 1,293,103 & 1,283,614  \\ 
% \textbf{MOOC}   & 7,047 & 97 & 411,749  & 345,600  \\ 
% \textbf{Wikipedia}  & 8,227 & 1,000 & 157,474  & 152,757  \\ 
% \textbf{Yoochoose}  & 33,670 & 3,667 & 211,851  & 41,374 \\ 
% \textbf{Douban Movie} & 9,987  & 9,999  & 1,001,918  & 999,732  \\ 
% \textbf{Amazon-Video}  & 5,130 & 1,685 & 37,126  & 1,946 \\ 

% \bottomrule

% \end{tabular}
% }

% \vspace{-0.5cm}
% \end{table}

% \begin{table}[t]
% \footnotesize
% \centering
% \caption{Dataset statistics}
% \vspace{-0.2cm}
% \label{Table:dataset}
% \resizebox{.48\textwidth}{!}{
%  \renewcommand{\arraystretch}{1}
% \begin{tabular}{crrrc}
% \toprule
% \textbf{Datasets} & \textbf{\# Users} & \textbf{\# Items} & \textbf{\# Interactions} & \textbf{Edge Feature} \\ \midrule
% \textbf{LastFM} & 980  & 1,000  & 1,293,103 & X  \\ 
% \textbf{MOOC}   & 7,047 & 97 & 411,749  & O  \\ 
% \textbf{Wikipedia}  & 8,227 & 1,000 & 157,474  & O  \\ 
% \textbf{Yoochoose}  & 33,670 & 3,667 & 211,851  & X \\ 
% \textbf{Douban Movie} & 9,987  & 9,999  & 1,001,918  & X  \\ 
% \textbf{Amazon-Video}  & 5,130 & 1,685 & 37,126  & X \\ 

% \bottomrule

% \end{tabular}
% }

% \vspace{-0.5cm}
% \end{table}

%%%%%% baby, sports, office, game
\begin{table}[t]
    \centering
    \caption{Dataset statistics}
    \vspace{-3mm}
\scalebox{0.82}{
\begin{tabular}{c|cccc}
\toprule
\textbf{Dataset} & \textbf{\#Users} & \textbf{\#Items} & \textbf{\#Interactions} & \textbf{Sparsity} \\
\hline
\textbf{Baby}        & 19,445  & 7,050  & 160,792  & 99.8827 \\
\textbf{Sports}      & 35,598  & 18,357 & 296,337  & 99.9547 \\
% Elec      & 108,693 & 37,460 & 1,689,188 & 99.9585 \\
% Clothing  & 39,387  & 23,033 & 278,677  & 99.9693 \\
\textbf{Office}      & 4,905   & 2,420  & 53,258   & 99.5513 \\
\textbf{Video Games} & 24,303  & 10,672 & 231,780  & 99.9106 \\
% Grocery   & 14,681  & 8,713  & 151,254  & 99.8818 \\
\bottomrule
\end{tabular}}
\label{tab:dataset}
% \vspace{-7mm}
\end{table}

%%%%%% all
% \begin{table}[t]
%     \centering
%     \caption{Dataset Statistics}
%     \vspace{-3mm}
% \scalebox{0.8}{
% \begin{tabular}{c|cccc}
% \toprule
% Dataset & \#Users & \#Items & \#Inters. & Sparsity \\
% \hline
% Baby     & 19,445  & 7,050  & 160,792  & 99.8827 \\
% Sports   & 35,598  & 18,357 & 296,337  & 99.9547 \\
% Elec     & 108,693 & 37,460 & 1,689,188 & 99.9585 \\
% Clothing & 39,387  & 23,033 & 278,677  & 99.9693 \\
% Office   & 4,905   & 2,420  & 53,258   & 99.5513 \\
% Game     & 24,303  & 10,672 & 231,780  & 99.9106 \\
% Grocery  & 14,681  & 8,713  & 151,254  & 99.8818 \\
% \bottomrule
% \end{tabular}}
% \label{tab:dataset}
% \vspace{-5mm}
% \end{table}

% \begin{table}[t]
%     \centering
%     \caption{Dataset Statistics}
%     \vspace{-3mm}
% \scalebox{0.8}{
% \begin{tabular}{c|ccc|ccc}
% \toprule
% \multirow{2}{*}{Dataset} & \multicolumn{3}{c|}{Raw Data} & \multicolumn{3}{c}{5-core Data} \\ 
% \cline{2-7} 
% ~   &   \#Users    & \#Items      & \#Inters.      &  \#Users     &  \#Items     &   \#Inters.   \\ 
% \hline
% Baby   &   531,890    &    71,317   &   915,446    &  19,445     &  7,050    & 160,792     \\
% Sports   &   1,990,521    &    532,197   &    3,268,695   &  35,598     &  18,357     &  296,337    \\
% Office & 
% Video Games & 
% % Electronics     &   4,201,696    &   498,196   & 7,824,482  &   192,403    &  63,001     &   1,689,188      \\ 
% \bottomrule
% \end{tabular}}
% \label{tab:dataset}
% \vspace{-5mm}
% \end{table}

\subsection{Experimental Setup}

\noindent\textbf{Datasets.} 
We evaluate \ours\ on four Amazon datasets~\cite{he2016ups, mcauley2015image}: Baby, Sports, Office, and Video Games.
Each dataset contains user-item interactions along with textual descriptions and images for each item.
From these sources, we extract (1) 4,096-dimensional visual features using VGGNet~\cite{simonyan2014very}, 
(2) 384-dimensional textual features using BERT~\cite{devlin2019bert}, 
and (3) 768-dimensional multimodal features using BEiT3~\cite{wang2023image}.
Dataset statistics are summarized in \textbf{Table~\ref{tab:dataset}}.

\vspace{1mm}
\noindent\textbf{Baselines.}
We compare \ours\ against eight state-of-the-art MMRS methods: VBPR~\cite{he2016vbpr}, LATTICE~\cite{zhang2021mining}, FREEDOM~\cite{zhou2023tale}, 
LGMRec~\cite{guo2024lgmrec}, SMORE~\cite{ong2025spectrum},
BM3~\cite{zhou2023bootstrap}, 
DA-MRS~\cite{xv2024improving}, and AlignRec~\cite{liu2024alignrec}.

\begin{table*}[!t]
    \centering
    \footnotesize
    \caption{Overall top-$\boldsymbol{N}$ recommendation performance across four Amazon datasets}
    \vspace{-0.4cm}
\scalebox{0.95}{
\begin{tabular}{c|c| ccccc | cccc}
\toprule
\textbf{Dataset}  & \textbf{Metric}
    & \multicolumn{5}{c}{\textbf{Accuracy-oriented MMRSs}}
    & \multicolumn{4}{|c}{\textbf{Alignment-based MMRSs}} \\
\cmidrule(lr){3-7}\cmidrule(lr){8-11}
&
    & \textbf{VBPR} & \textbf{LATTICE} & \textbf{FREEDOM} & \textbf{LGMRec} & \textbf{SMORE}
    & \textbf{BM3} & \textbf{DA-MRS} & \textbf{AlignRec} & \textbf{AnchorRec} \\ \hline
\multirow{2}{*}{\textbf{Baby}} 
    &R@20  & 0.0600 & 0.0859 & 0.0976 & 0.0980 & \underline{0.0997} & 0.0831 & 0.0947 & 0.1007 & \textbf{\underline{0.1013}} \\
    &N@20  & 0.0261 & 0.0375 & 0.0420 & 0.0436 & \underline{0.0443} & 0.0357 & 0.0420 & \textbf{\underline{0.0444}} & 0.0436 \\
\hline
\multirow{2}{*}{\textbf{Sports}} 
    &R@20  & 0.0730 & 0.0950 & 0.1034 & 0.1068 & \textbf{\underline{0.1135}} & 0.0907 & 0.1073 & \textbf{\underline{0.1135}} & 0.1048\\
    &N@20  & 0.0329 & 0.0424 & 0.0455 & 0.0479 & \textbf{\underline{0.0508}} & 0.0400 & 0.0480 & \underline{0.0507} & 0.0458 \\
\hline
\multirow{2}{*}{\textbf{Office}} 
    &R@20  & 0.1069 & 0.1385 & 0.1466 & 0.1474 & \textbf{\underline{0.1619}} & 0.1144 & 0.1475 & 0.1446 & \underline{0.1513}\\
    &N@20  & 0.0505 & 0.0636 & 0.0685 & 0.0690 & \textbf{\underline{0.0773}} & 0.0532 & \underline{0.0701} & 0.0656 & 0.0694\\
\hline
\multirow{2}{*}{\textbf{Video Games}} 
    &R@20  & 0.1681 & 0.1727 & 0.1651 & \textbf{\underline{0.1990}} & 0.1919 & 0.1651 & \underline{0.1941} & 0.1847 & 0.1805\\
    &N@20  & 0.0750 & 0.0769 & 0.0726 & \textbf{\underline{0.0900}} & 0.0858 & 0.0728 & \underline{0.0866} & 0.0827 & 0.0790\\
\bottomrule
\end{tabular}}
\label{tab:overall}
\vspace{-0mm}
\end{table*}

\vspace{1mm}
\noindent\textbf{Implementation Details.}
We follow a standard MMRS evaluation protocol, splitting each dataset into training, validation, and test sets with an 8:1:1 ratio.
Top-$N$ recommendation performance is measured using Recall@$N$~\cite{powers2020evaluation} and NDCG@$N$~\cite{jarvelin2002cumulated}, with results averaged over \textit{five} random seeds.
All baselines and \ours\ are implemented using the official MMRec framework~\cite{zhou2023mmrec}, and hyperparameters are tuned via grid search on the validation set.

\subsection{Results}

\vspace{1mm}
\noindent{\bf (1) Top-$\boldsymbol{N}$ Recommendation Performance.}
% our goal is to enhance multimodal expressiveness while sacrificing as little accuracy as possible.
\textbf{Table~\ref{tab:overall}} reports top-$N$ recommendation results across datasets.
Overall, accuracy-oriented MMRSs such as LGMRec and SMORE achieve the strongest accuracy, reflecting their focus on ranking optimization.
Despite not pursuing aggressive ranking objectives, \ours\ remains competitive and often matches or surpasses existing alignment-based methods (BM3, DA-MRS, and AlignRec) on multiple datasets and metrics.
Importantly, our goal is to enhance multimodal expressiveness while sacrificing as little accuracy as possible, which we further validate in the following analyses.

\begin{figure}[t]
\centering
\includegraphics[width=\linewidth]{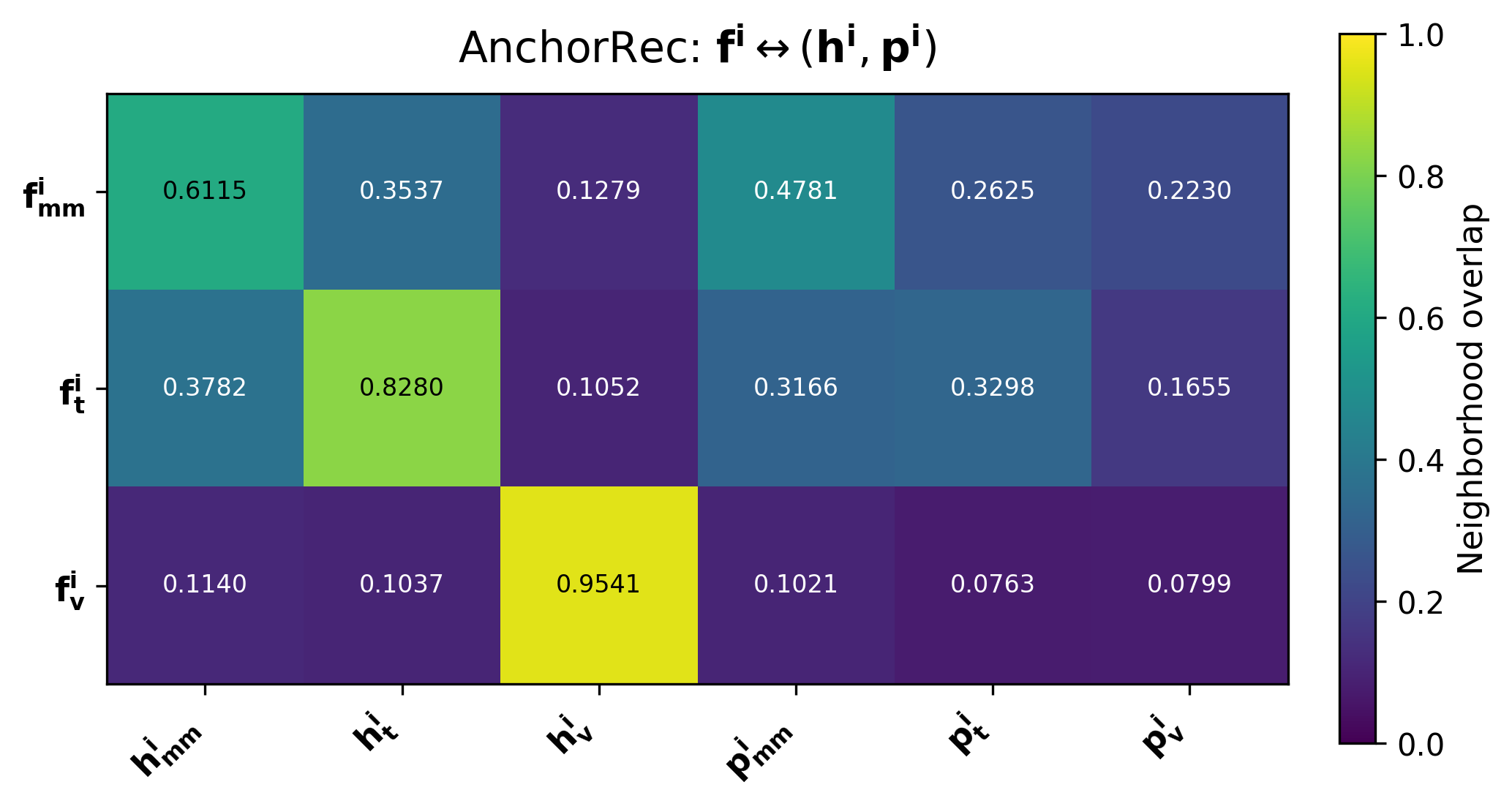}
\vspace{-0.6cm}
\caption{Neighborhood overlap across embedding spaces.} \label{fig:RQ3-1}
\vspace{-0cm}
\end{figure}

\vspace{1mm}
\noindent{\bf (2) Preserving Modality-specific Structure.}
% We investigate whether \ours\ simultaneously (a) aligns modalities coherently and (b) preserves their distinctive semantics.
We examine whether \ours\ can achieve cross-modal alignment while preserving modality-specific structural properties.
% simultaneously (a) aligns modalities coherently and (b) preserves modality-specific structural characteristics.
For each item, we construct \textit{Top-$K$ neighborhood} based on cosine similarity for every modality and measure neighborhood overlap across modalities.
% A \textit{high} overlap means the modalities behave similarly,
% whereas a \textit{low} overlap shows each modality keeps its own characteristics.
\textit{Higher} overlap indicates shared  neighborhood structure, whereas \textit{lower} overlap suggests that modality-specific characteristics are preserved.
% ver1
% \textbf{Figure~\ref{fig:RQ3-1}} visualizes the similarity-overlap matrices across modalities to examine their structural relationships in different representation spaces.
% Specifically, \textbf{Figure~\ref{fig:RQ3-1}-(a)} compares the overlap among raw pre-trained features $(\mathbf{f}_{mm}^i, \mathbf{f}_t^i, \mathbf{f}_v^i)$, the corresponding origin embeddings $(\mathbf{h}_{mm}^i, \mathbf{h}_t^i, \mathbf{h}_v^i)$, and the projection-domain embeddings $(\mathbf{p}_{mm}^i, \mathbf{p}_t^i, \mathbf{p}_v^i)$ in \ours.
% \textbf{Figure~\ref{fig:RQ3-1}-(b)} presents the similarity-overlap comparison between raw pre-trained features and origin embeddings for a representative baseline MMRS, AlignRec.
% ver2
\textbf{Figure~\ref{fig:RQ3-1}} presents the neighborhood overlap matrix across different embedding spaces: (1) raw pre-trained features (\ie, $\textbf{f}_{mm}^i, \textbf{f}_{t}^i,$ and $\textbf{f}_{v}^i$), (2) original-space embeddings (\ie, $\textbf{h}_{mm}, \textbf{h}_{t}^i,$ and $\textbf{h}_{v}^i$), and (3) projection-domain embeddings (\ie, $\textbf{p}_{mm}^i, \textbf{p}_{t}^i,$ and $\textbf{p}_{v}^i$).
% and (4) final fused embeddings (\ie, $\textbf{h}^i$).

% ver1
% We observe two key findings:
% In the \textbf{origin embedding space and raw feature space} of Figure~\ref{fig:RQ3-1}-(a), a high overlap is observed between neighborhood sets of the same modality pairs, indicating that the modality-preserving objective in AMP successfully retains modality-specific representational characteristics even after the alignment process.
% In contrast, in the \textbf{projection domain}, neighborhood structures across different modalities exhibit reduced distinction, suggesting that the anchor-based alignment strategy in AAL, which uses the multimodal representation as the anchor, prioritizes cross-modal structural alignment rather than preserving modality-specific neighborhood structures.
% Furthermore, Figure~\ref{fig:RQ3-1}-(b) shows that, due to the direct alignment strategy adopted by AlignRec, modality-specific characteristics are diminished, and the resulting embedding relationships become similar to those observed in the projection space of \ours.
% ver2
We observe two key findings:
First, when comparing raw pretrained features and origin embeddings, each modality largely preserves its original neighborhood structure, as evidenced by high diagonal overlap and low cross-modality overlap.
This indicates that the learned origin embeddings retain inherent modality-specific properties.
Second, in the projection domain, neighborhood overlap with raw features becomes uniformly low, showing that projection-domain embeddings intentionally deviate from original modality structures to facilitate alignment.
Notably, projected embeddings become more similar to each other via the multimodal anchor, confirming that alignment is achieved in the projection space while preserving raw modality structure in the origin embeddings.

% Together, these results show that \ours\ separates alignment from representation learning, enabling coherent multimodal alignment while preserving modality-specific expressiveness.

% demonstrating effective cross-modal consistency, which indicates that the anchor-based alignment strategy in AAL, using the multimodal representation as the anchor, successfully achieves structural alignment across modalities.
% In the \textbf{origin and raw spaces}, neighborhood structures remain distinct across modalities, confirming that the modality-preserving objective in AMP effectively retains original modality-specific properties even after the alignment process.
% These behaviors confirm that \ours\ preserves modality-specific expressiveness without overfitting to any single modality, achieving a balanced alignment across modalities.

\begin{figure}[t]
\centering
\includegraphics[width=0.94\linewidth]{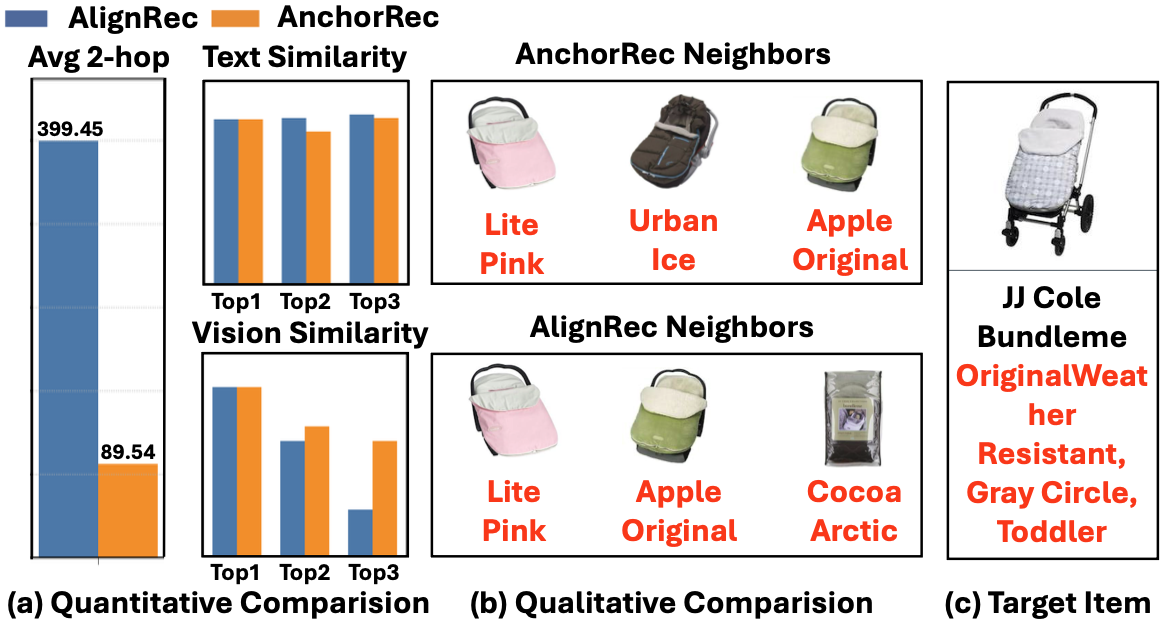}
\vspace{-0.3cm}
\caption{
Top-3 neighbors retrieved by AlignRec and \ours\ for the same target item:
(a) quantitative comparison using 2-hop proximity and text and vision similarity, and
(b) qualitative comparison using images and textual keywords.
% Top-3 neighbors retrieved by AlignRec and AnchorRec using the same target item. (a) Quantitative comparison based on the number of 2-hop items and the cosine similarity of image and text embeddings. (b) Qualitative comparison using the image and text metadata of the retrieved items.
}\label{fig:RQ2}
\vspace{0cm}
\end{figure}

\vspace{1mm}
\noindent{\bf (3) Expressiveness of Final Item Embedding.}
We examine wheth\ er the final fused embeddings preserve expressive signals from individual modalities.
\textbf{Figure~\ref{fig:RQ2}} compares the top-3 nearest neighbors retrieved by AlignRec and \ours\ for the same target item, shown on the right of \textbf{Figure~\ref{fig:RQ2}-(c)}.
\textbf{Figure~\ref{fig:RQ2}-(a)} reports the average 2-hop proximity and the visual and textual cosine similarity between the target item and its neighbors.
AlignRec tends to favor items with strong interaction proximity, as evidenced by substantially higher average 2-hop values, even when visual or textual similarity is low. %resulting in retrieved neighbors with higher average interaction scores.
In contrast, \ours\ retrieves neighbors that balance interaction-based relevance with visual and textual similarity.
\textbf{Figure~\ref{fig:RQ2}-(b)} further illustrates this difference qualitatively.
Neighbors retrieved by \ours\ exhibit clearer semantic similarity to the target item (\eg, color, style, and usage), whereas AlignRec often retrieves interaction-related but semantically less aligned items (\eg, the third retrieved neighbor).
% This pattern is consistently observed across different product categories.

% For each neighbor, we report user-behavior proximity (\textit{2-hop}), visual similarity (\textit{img cos}), and textual similarity (\textit{txt cos}), together with the corresponding product image and raw description.
% As shown in \textbf{Figure~\ref{fig:RQ2}-(a)}, AlignRec tends to favor items with strong interaction proximity, as indicated by higher average 2-hop values, even when visual or textual similarity is low.
% In contrast, \ours\ retrieves neighbors with a more balanced combination of interaction-based, visual, and textual relevance.
% This is further reflected in \textbf{Figure~\ref{fig:RQ2}-(b)}, where neighbors retrieved by \ours\ exhibit clearer semantic similarity to the target item (\eg, color, style, and usage), while AlignRec often retrieves interaction-related but semantically less aligned items, as illustrated by the third retrieved neighbor.
% We observe consistent trends across different product categories.

% dissimilar to the target item, which is further reflected by the high average interaction scores shown in \textbf{Figure~\ref{fig:RQ2}-(a)}.
% In contrast, \textbf{\ours} retrieves neighbors that exhibit a more \textit{balanced combination} of visual, textual, and interaction-based relevance.
% % modify
% Moreover, the qualitative comparison in \textbf{Figure~\ref{fig:RQ2}-(b)} further confirms that \ours\ retrieves semantically more similar items. 
% % modify
% This trend remains consistent across different product categories. 

\vspace{1mm}
\noindent{\bf (4) Supplementary Analyses.}
Additional analyses are provided at \href{https://sites.google.com/view/anchorrec-preprint}{\url{https://sites.google.com/view/anchorrec-preprint}}, including ablation studies (\eg, w/o $\mathcal{L}_{\mathrm{anchor}}$ and w/o projection), hyperparameter sensitivity (\eg, $\lambda_{1}$, $\lambda_{2}$, and the InfoNCE temperature $\tau$), and computational efficiency analysis.

% \section{Evaluation} \label{sec:eval} \input{Sections/short/Eval_short}

\section{Conclusion} \label{sec:concl} In this work, we identified a key limitation of current MMRSs: direct alignment in a unified space collapses modality-specific representations and amplifies ID dominance.
To address this issue, we proposed \ours, a projection-domain alignment framework that enables anchor-based indirect alignment, preserving modality-specific expressiveness while achieving cross-modal consistency.
Experiments on four Amazon datasets demonstrate that \ours\ maintains competitive recommendation accuracy while significantly enhancing multimodal expressiveness.

\bibliographystyle{ACM-Reference-Format}
\bibliography{bibliography}

%\appendix \label{sec:app} \input{Appendix}
% \clearpage
\setcounter{table}{0}
\setcounter{figure}{0}

% \textbf{Appendix}
\label{sec:appendix}% \appendix

\end{document}